\def\papertitle{Clean2FX: Label-Conditioned Modeling for Clean-to-Effect Guitar Audio Transformations}
\def\paperauthorA{Oliverio Bombicci Pontelli}
\def\paperauthorB{Iran R. Roman}

\documentclass[twoside,a4paper]{article}
\usepackage{etoolbox}

\usepackage[print]{dafx26demo}

\usepackage{amsmath,amssymb,amsfonts,amsthm}
\usepackage{siunitx}
\usepackage{euscript}
\usepackage[T1]{fontenc}
\usepackage[utf8]{inputenc}
\usepackage{ifpdf}
\usepackage[english]{babel}
\usepackage{caption}
\usepackage{subfig}
\usepackage{color}
\usepackage{xcolor}
\usepackage{booktabs}
\usepackage{makecell}

\newcommand{\better}[1]{\textcolor[HTML]{1D9E75}{#1}}
\newcommand{\worse}[1]{\textcolor[HTML]{D85A30}{#1}}
\newcommand{\pos}[1]{\textcolor[HTML]{1D9E75}{#1}}
\newcommand{\nega}[1]{\textcolor[HTML]{D85A30}{#1}}

\input glyphtounicode
\ninept
\newcounter{numauth}
\newcounter{listcnt}
\newcommand\authcnt[1]{\ifdefined#1 \stepcounter{numauth} \fi}

\newcommand\addauth[1]{
\ifdefined#1
\stepcounter{listcnt}
\ifnum \value{listcnt}<\value{numauth}
\appto\authorslist{, #1}
\else
\appto\authorslist{~and~#1}
\fi
\fi}
\authcnt{\paperauthorB}
\authcnt{\paperauthorC}
\authcnt{\paperauthorD}
\authcnt{\paperauthorE}
\authcnt{\paperauthorF}
\authcnt{\paperauthorG}
\authcnt{\paperauthorH}
\authcnt{\paperauthorI}
\authcnt{\paperauthorJ}
\def\authorslist{\paperauthorA}
\addauth{\paperauthorB}
\addauth{\paperauthorC}
\addauth{\paperauthorD}
\addauth{\paperauthorE}
\addauth{\paperauthorF}
\addauth{\paperauthorG}
\addauth{\paperauthorH}
\addauth{\paperauthorI}
\addauth{\paperauthorJ}

\usepackage{times}

\newif\ifpdf
\ifx\pdfoutput\relax
\else
   \ifcase\pdfoutput
      \pdffalse
   \else
      \pdftrue
   \fi
\fi

\ifpdf 
  \usepackage[pdftex,
    pdftitle={\papertitle},
    pdfauthor={\authorslist},
    pdfsubject={Proceedings of the 29th International Conference on Digital Audio Effects (DAFx26)},
    colorlinks=false,
    bookmarksnumbered,
    pdfstartview=XYZ
  ]{hyperref}
  \usepackage[pdftex]{graphicx}
\else 
  \usepackage[dvips]{epsfig,graphicx}
  \usepackage[dvips,
    pdftitle={\papertitle},
    pdfauthor={\authorslist},
    pdfsubject={Proceedings of the 29th International Conference on Digital Audio Effects (DAFx26)},
    colorlinks=false,
    bookmarksnumbered,
    pdfstartview=XYZ
  ]{hyperref}
\fi
\usepackage[hypcap=true]{caption}
\title{\papertitle}

\affiliation
{
\begin{minipage}[t]{0.45\textwidth}
\centering
\paperauthorA\\
\href{https://www.qmul.ac.uk/eecs/}{School of Electronic Engineering \& Computer Science}\\
Queen Mary University of London\\
London, United Kingdom\\
{\tt \href{mailto:oliveriobp@gmail.com}{oliveriobp@gmail.com}}
\end{minipage}
\hfill
\begin{minipage}[t]{0.45\textwidth}
\centering
\paperauthorB\\
\href{https://www.qmul.ac.uk/eecs/}{School of Electronic Engineering \& Computer Science}\\
Queen Mary University of London\\
London, United Kingdom\\
{\tt \href{mailto:i.roman@qmul.ac.uk}{i.roman@qmul.ac.uk}}
\end{minipage}
}
{}

   \let\OLDthebibliography\thebibliography
   \renewcommand\thebibliography[1]{%
     \OLDthebibliography{#1}%
     \setlength{\parskip}{0pt}%
     \setlength{\itemsep}{0.9pt plus 0.3ex}%
   }

\begin{document}
\ifpdf
  \DeclareGraphicsExtensions{.png,.jpg,.pdf}
\else
  \DeclareGraphicsExtensions{.eps}
\fi

\maketitle

\begin{abstract} We present Clean2FX, a study and demo of label-conditioned clean-to-effect transformation for electric guitar audio. Given a clean guitar input and a target effect label, the task is to synthesize the corresponding effected signal while preserving the musical content. Training and evaluation pairs are constructed from EGFxSet real, single-tone recordings by assembling matched clean/effected chords, melodies, and mixed timelines. This allows for controlled comparison across effects. We evaluate four neural approaches under a common spectrogram-based transformation setting: two variational autoencoders and two U-Net models that differ in whether they operate on linear or log-magnitude representations. Performance is measured using linear-magnitude spectrogram MSE and Fr\'{e}chet Audio Distance. The U-Net models outperform the variational autoencoder variants. Per-effect results show that distortion effects are most readily improved, whereas delay and reverb effects exhibit weaker FAD gains despite substantial spectral-error reductions. A conditioning-sensitivity diagnostic provides evidence that the best model responds to target labels rather than collapsing to a single transformation. Our demo website compares two models applied on real-world guitar performances outside training and validation data, providing audio and spectrogram examples of the practical clean-to-effect behavior. \end{abstract} 

\section{Introduction} \label{sec:intro}  Effects are central to the sound of the electric guitar: they reshape timbre, dynamics, sustain, and apparent space, and have become part of both the instrument's musical vocabulary and the broader history of audio effects processing~\cite{hunter_guitar_effects, ostberg_electric_guitar,schneider_contemporary_guitar,wilmering_history}. From a signal-processing perspective, effects can be understood as transformations of a clean instrumental source~\cite{smith2007introduction}.  Recent neural approaches to audio effects have shown that input--output behavior can be learned directly from paired recordings, including black-box modeling of audio effects~\cite{mr_blackbox}, real-time guitar-amplifier emulation~\cite{wright_amp}, and controllable audio-effect transformation~\cite{deepafx}. Much of this work, however, is organized around a single device, a single effect class, or a parametric processor. 
Clean-to-effect guitar performance transformation across several effect families poses a problem: preservation of pitch, rhythm, and playing content while applying a requested transformation whose acoustic footprint may range from broadband distortion to modulation, delay, or reverb.  

This paper presents Clean2FX as both a study of label-conditioned clean-to-effect guitar audio transformation and a web demo. We construct paired clean/effected examples from EGFxSet, a dataset of electric-guitar tones processed through real hardware effects~\cite{egfxset,Pedroza2024Leveraging}, by assembling matched chords, melodies, and mixed timelines. This gives a controlled setting in which the musical content is held fixed and the target effect label defines the transformation. We compare four neural approaches: two variational autoencoders and two conditional encoder--decoder U-Nets. 

The demo website complements this controlled evaluation by applying two trained systems to real-world guitar performances~\cite{guitartechs} outside the EGFxSet-derived material. The demo therefore illustrates the practical behavior of the models beyond synthetic data.\footnote{\url{guitar-clean2fx-demo.netlify.app}} The code implementation is released too.\footnote{\url{github.com/oliveriobp/fyp-clean2fx}}

\section{Methods}
\label{sec:methods}

We build our training data using EGFxSet~\cite{egfxset}, which contains isolated electric guitar tones recorded through ten real hardware effects alongside a clean reference signal. We want to train models on musical sequences, so we create synthetic performances programmatically, using these recorded tone sounds. Our resulting custom dataset assembles chords, melodies, and mixed event timelines, applying the same event sequence to both the clean and effected versions of each tone. This pairing holds all musical content constant across both signals, leaving the effect as the sole axis of variation in each training example. Each pair is normalized by the larger of its two signal maxima to preserve the energy ratio between the clean and effected signals. The probability of sampling near-silent compositions is set to 0.05.

Audio is represented as magnitude STFT spectrograms with $n_{\text{fft}}=512$ and hop $h=128$ at \SI{16}{kHz}; each \SI{4.9}{s} clip yields a $257\times613$ spectrogram. Two input representations are compared: the linear shared-max normalized magnitude $x$, and a log-compressed variant $y=\ln(1+200x)$, inverted as $x=(e^{y}-1)/200$ before any metric is computed. Since only magnitude is retained, the phase of the output spectrogram is unavailable; at inference, it is estimated using Griffin-Lim~\cite{griffinlim}.

\begin{figure*}[t]
  \centering
  \includegraphics[width=0.9\textwidth]{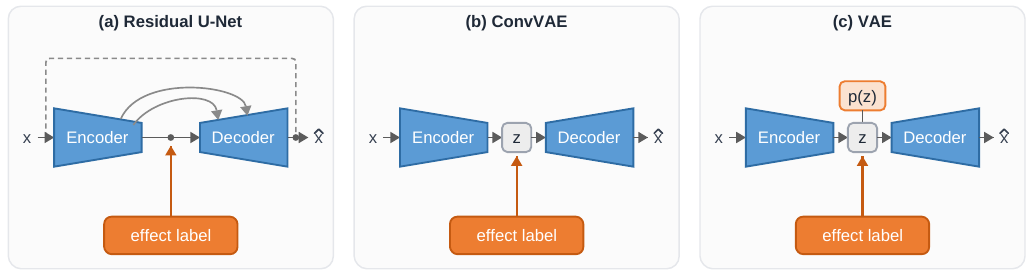}
  \caption{The three model families compared in this work. The U-Net and ConvVAE have a convolutional encoder--decoder structure, conditioned on the target effect label.
  (a)~Residual U-Net: skip connections link encoder and decoder, FiLM modulates the decoder levels and the skip connections, and a global residual path adds the clean input. 
  (b)~ConvVAE: a deterministic latent $z$ and no skip connections. (c)~VAE: uses dense layers, and the latent $z$ is sampled from $p(z)$. $x$ and $\hat{x}$ are the input and output magnitude spectrograms, respectively.}
  \label{fig:models}
\end{figure*}

\begin{table}[t]
\centering
\caption{Evaluation of models conditioned to transform clean electric guitar recordings
into target effects from EGFxSet~\cite{egfxset} (a dataset of real
hardware-processed guitar tones). The unprocessed input baseline measures
the error when comparing against the clean recording unchanged (i.e., applying no
effect) and defines the minimum bar a useful model must exceed. 
The two variants differ only in
whether the loss is applied to linear~($\mathcal{L}_{\text{lin}}$) or
log-magnitude~($\mathcal{L}_{\text{log}}$) spectra. MSE: mean squared
error on linear shared-max magnitude spectrograms ($\times10^{-4}$; $\downarrow$ better).
FAD: Fr\'{e}chet Audio Distance, a perceptual quality metric
($\downarrow$ better). $\Delta$: mean per-effect improvement over the
unprocessed input baseline ($\uparrow$ better); ``no imp'' indicates the model is,
on average across effects, worse than applying no effect.}
\label{tab:results}
\footnotesize
\begin{tabular}{lcccc}
\toprule
& \multicolumn{2}{c}{$\downarrow$ MSE$_ \text{($\times10^{-4}$)}$}
& \multicolumn{2}{c}{$\downarrow$ FAD} \\
\cmidrule(lr){2-3} \cmidrule(lr){4-5}
Model & Value & $\Delta$ & Value & $\Delta$ \\
\midrule
\multicolumn{5}{l}{\textit{Baseline}} \\[2pt]
\quad Unprocessed input        & 5.16  & ---           & 6.82  & ---           \\
\midrule
\multicolumn{5}{l}{\textit{Learned models}} \\[2pt]
\quad VAE                      & 11.44 & \nega{no imp} & 17.53 & \nega{no imp} \\
\quad ConvVAE                  & 4.09  & \nega{no imp} & 7.38  & \nega{no imp} \\
\quad U-Net ($\mathcal{L}_{\text{lin}}$) & 1.08  & \pos{$+$69.4\%} & 3.98  & \pos{$+$23.0\%} \\
\quad U-Net ($\mathcal{L}_{\text{log}}$) & \textbf{1.07} & \pos{\textbf{$+$70.6\%}} & \textbf{3.45} & \pos{\textbf{$+$31.8\%}} \\
\bottomrule
\end{tabular}
\end{table}

\begin{table*}[t]
\centering
\caption{Per-effect breakdown of MSE and FAD improvements corresponding to
Table~\ref{tab:results}, grouped by effect category.
Unprocessed input columns show raw baseline values for each
metric (lower is better); note that RAT has a baseline MSE an order of
magnitude larger than all other effects, making it the easiest target;
a pooled aggregate would be dominated by RAT, which is why
Table~\ref{tab:results} reports improvement averaged over effects.
$\Delta$MSE and $\Delta$FAD give percentage improvement over the
unprocessed input baseline ($\uparrow$ better);
"\worse{no imp}"~indicates performance below baseline.
VAE is omitted: apart from a marginal $+$0.7\% on RAT by MSE, it shows
no improvement on any effect on either metric.
Delay effects show notably weaker FAD gains despite strong MSE
improvements, suggesting spectral error and perceptual quality diverge
for time-based effects.
Bold marks the best $\Delta$MSE and the best $\Delta$FAD in each row among the models.}
\label{tab:per-effect}
\footnotesize
\setlength{\tabcolsep}{5pt}
\begin{tabular}{l rr rr rr rr}
\toprule
& \multicolumn{2}{c}{Unprocessed input}
& \multicolumn{2}{c}{ConvVAE}
& \multicolumn{2}{c}{U-Net ($\mathcal{L}_{\text{lin}}$)}
& \multicolumn{2}{c}{U-Net ($\mathcal{L}_{\text{log}}$)} \\
\cmidrule(lr){2-3}\cmidrule(lr){4-5}
\cmidrule(lr){6-7}\cmidrule(lr){8-9}
Effect
  & \makecell{$\downarrow$ MSE$_\text{\scalebox{.75}{$\times10^{-4}$}}$}
  & \makecell{$\downarrow$ FAD}
  & $\Delta$MSE & $\Delta$FAD
  & $\Delta$MSE & $\Delta$FAD
  & $\Delta$MSE & $\Delta$FAD \\
\midrule
\multicolumn{9}{l}{\textit{Distortion}} \\[2pt]
\quad RAT
  & 32.16 & 23.84
  & \better{$+$65.9\%} & \better{$+$64.3\%}
  & \better{\textbf{$+$85.0\%}} & \better{$+$78.3\%}
  & \better{$+$84.8\%} & \better{\textbf{$+$82.5\%}} \\
\quad Tube Screamer
  & 3.42 & 10.96
  & \worse{no imp} & \better{$+$50.9\%}
  & \better{\textbf{$+$82.8\%}} & \better{$+$55.4\%}
  & \better{$+$80.0\%} & \better{\textbf{$+$57.0\%}} \\
\midrule
\multicolumn{9}{l}{\textit{Modulation}} \\[2pt]
\quad Flanger
  & 3.39 & 9.56
  & \worse{no imp} & \worse{no imp}
  & \better{$+$56.7\%} & \better{$+$16.1\%}
  & \better{\textbf{$+$57.2\%}} & \better{\textbf{$+$28.3\%}} \\
\quad Phaser
  & 1.41 & 4.06
  & \worse{no imp} & \worse{no imp}
  & \better{$+$66.9\%} & \better{$+$21.7\%}
  & \better{\textbf{$+$71.1\%}} & \better{\textbf{$+$44.9\%}} \\
\midrule
\multicolumn{9}{l}{\textit{Delay}} \\[2pt]
\quad Digital Delay
  & 0.65 & 1.40
  & \worse{no imp} & \worse{no imp}
  & \better{\textbf{$+$50.9\%}} & \worse{no imp}
  & \better{$+$49.9\%} & \better{\textbf{$+$0.9\%}} \\
\quad Sweep Echo
  & 3.14 & 1.21
  & \worse{no imp} & \worse{no imp}
  & \better{\textbf{$+$64.0\%}} & \better{$+$25.9\%}
  & \better{$+$63.5\%} & \better{\textbf{$+$33.4\%}} \\
\quad Tape Echo
  & 1.63 & 2.08
  & \worse{no imp} & \worse{no imp}
  & \better{$+$68.0\%} & \better{\textbf{$+$25.7\%}}
  & \better{\textbf{$+$74.0\%}} & \better{$+$23.7\%} \\
\midrule
\multicolumn{9}{l}{\textit{Reverb}} \\[2pt]
\quad Hall Reverb
  & 1.90 & 6.54
  & \worse{no imp} & \worse{no imp}
  & \better{$+$74.7\%} & \better{$+$13.8\%}
  & \better{\textbf{$+$76.4\%}} & \better{\textbf{$+$24.2\%}} \\
\quad Plate Reverb
  & 1.20 & 3.29
  & \worse{no imp} & \worse{no imp}
  & \better{$+$69.9\%} & \better{$+$11.5\%}
  & \better{\textbf{$+$70.3\%}} & \better{\textbf{$+$23.1\%}} \\
\quad Spring Reverb
  & 2.71 & 5.26
  & \worse{no imp} & \worse{no imp}
  & \better{$+$75.4\%} & \worse{no imp}
  & \better{\textbf{$+$78.6\%}} & \better{\textbf{$+$0.9\%}} \\
\midrule
\multicolumn{9}{l}{\textit{Mean across effects}} \\[2pt]
\quad All effects
  & 5.16 & 6.82
  & \worse{no imp} & \worse{no imp}
  & \better{$+$69.4\%} & \better{$+$23.0\%}
  & \better{\textbf{$+$70.6\%}} & \better{\textbf{$+$31.8\%}} \\
\bottomrule
\end{tabular}
\end{table*}

We assess four architectures, each conditioned on a learned effect embedding. Figure~\ref{fig:models} summarizes and illustrates them. The VAE encodes the input with fully connected layers into a 64-dimensional latent and decodes symmetrically, trained with the standard VAE objective (MSE reconstruction plus a KL term). The ConvVAE replaces the dense layers with strided convolutions using BatchNorm and ReLU, a 512-dimensional deterministic latent, and an L1 loss on $\log(1+x)$ magnitudes. We compare these VAEs against a five-level residual U-Net~\cite{unet,unet_spec}. The encoder comprises four DoubleConv blocks (channels 32, 64, 128, 256); each two $3\times3$ convolutions with Group Normalization (eight groups~\cite{groupnorm}) and LeakyReLU (slope 0.01), separated by $2\times2$ max-pooling; the bottleneck adds a fifth block at 512 channels; the decoder mirrors the encoder with nearest-neighbor upsampling and skip connections. The network predicts a residual summed with the input~\cite{resnet}, passed through softplus ($\beta=5$ for linear inputs, $\beta=1$ for log) to produce non-negative magnitudes. Two U-Net variants are evaluated, differing only in whether the loss operates on linear~($\mathcal{L}_{\text{lin}}$) or log-magnitude~($\mathcal{L}_{\text{log}}$) spectrograms. Throughout, we refer to a copy-input baseline, defined as the trivial system that outputs the clean input unchanged; this baseline anchors all relative-improvement figures reported in evaluation.

In the U-Net, each of the ten effect labels receives a learned 64-dimensional embedding, injected at three points. At the bottleneck the embedding is projected to 64 channels, broadcast, concatenated, and fused by a DoubleConv. At each of the four decoder levels, feature-wise linear modulation (FiLM)~\cite{film} maps the embedding to per-channel scale and shift parameters applied as $x\cdot(1+\gamma)+\beta$, where the $(1+\gamma)$ form is the identity at initialization; FiLM transfers cleanly to U-Nets~\cite{cunet}. Each encoder skip connection receives FiLM of the same form through its own per-level heads, so no path through the network bypasses the condition. Conditioning dropout zeroes the entire embedding with probability 0.15 during training, inspired by classifier-free guidance~\cite{ho2021classifier}, so the network cannot learn to ignore its condition. The two VAE baselines condition on the target-effect index through a separate four-dimensional learned embedding, concatenated to the latent code before decoding (64-dimensional in the fully connected VAE, 512-dimensional in the ConvVAE). Conditioning therefore enters at a single point, with no FiLM, skip-connection modulation, or conditioning dropout.

The U-Net variants optimize a composite spectral loss whose base term is spectral convergence (SC), $\|\hat{Y}-Y\|_F/(\|Y\|_F+1.0)$, where the large epsilon of 1.0 keeps the ratio bounded on near-silent targets. For the log-magnitude variant, the loss is $\text{SC} + 10\,\text{L1}_{\log}$, with the L1 term computed directly in the network's native log-magnitude domain. For the linear variant, the loss is $\text{SC} + 10\,\text{L1}_{\text{lin}} + 5\,\text{L1}_{\log}$: an L1 term in the native linear-magnitude domain, plus an auxiliary L1 term computed on a log-compressed version of the same prediction, included to penalize errors in low-energy regions that the linear term underweights. A per-effect weight $w=\min((\Delta_{\max}/\Delta_e)^{0.5},\,5.0)$ is applied to each sample's loss based on its target effect, where $\Delta_e$ is the mean clean-to-effect L1 distance of that effect and $\Delta_{\max}$ the largest such value across effects; this upweights effects with naturally small clean-to-effect spectral differences, which would otherwise contribute a vanishingly small gradient signal. These objectives apply to the U-Net variants; the VAE objectives are given above.

We train the U-Nets with AdamW~\cite{adamw} (learning rate $3\times10^{-4}$, weight decay $1\times10^{-4}$), ReduceLROnPlateau (factor 0.5, patience 8, minimum $1\times10^{-6}$), gradient clipping at norm 1.0, batch size 32, a RandomSampler with replacement over 8000-sample epochs, and a 90/10 train/validation split. The VAE baselines are trained on the same data split, using Adam at $1\times10^{-3}$ with mixed precision; the full pipeline is built on librosa~\cite{librosa} and PyTorch~\cite{pytorch}.

We evaluate the four trained models on 200 held-out paired samples (20 per effect), drawn from the same split. All metrics are computed on linear shared-max magnitude spectrograms. Per-bin MSE is complemented by Fr\'{e}chet Audio Distance (FAD)~\cite{fad}. Baseline-relative improvement is $(b-m)/b$ for model value $m$ against the copy-input baseline $b$, and the overall value reported is the mean of the ten per-effect means.

\section{Results}
\label{sec:results}

Table~\ref{tab:results} shows that the U-Net (Log) is best on both metrics ($+$70.6\% MSE, $+$31.8\% FAD), the U-Net (Linear) is second ($+$69.4\% and $+$23.0\%), and both variational autoencoder baselines fail to beat the copy-input baseline on average, with no positive mean per-effect improvement on either metric.

Table~\ref{tab:per-effect} groups the effects into distortion, modulation, delay, and reverb. The U-Net (Log) is the strongest model in most cells, performing best for six effects by MSE and nine by FAD; the U-Net (Linear) takes the four remaining MSE cells and the single remaining FAD cell. The two distortions sit at the top of the baseline-relative ranking and the time-based effects at the bottom, an ordering that tracks the size of the copy-input baseline: a distortion leaves little of the clean signal intact and so leaves the most error to remove, whereas a delay leaves the input largely unchanged. The two metrics diverge most on these time-based effects, which post strong MSE improvements but markedly weaker FAD gains; under the U-Net (Log), for example, Spring Reverb improves $+$78.6\% on MSE but only $+$0.9\% on FAD. 

\section{Discussion}
\label{sec:discussion}

The strongest pattern in Table~\ref{tab:per-effect} is architectural since the U-Net's skip connections pass each encoder feature map directly to the corresponding decoder level. This likely preserves harmonic structure at every resolution, while the VAE family compresses the input through a latent bottleneck and loses fine detail. This is a previously documented source of blurred VAE reconstructions~\cite{larsen2016autoencoding}. RAT is the exception, the only case where ConvVAE clears the baseline on both metrics. The copy-input baseline is the weakest, so capturing the broad spectral envelope is enough to beat it.

We use effect macro-averaging because MSE on a magnitude spectrogram is scale-dependent: RAT's copy-input baseline is an order of magnitude above every other effect, so it looks hardest by absolute MSE yet is the easiest in relative terms, since the unchanged input leaves the most error to remove. Effect-averaging keeps it from dominating the aggregate: the ConvVAE, for instance, has a lower pooled MSE than the baseline ($4.09$ vs $5.16\times10^{-4}$) but clears the baseline on no effect except RAT, so its effect-averaged improvement is not positive.
Log compression helps because it lifts low-amplitude detail such as reverb tails, modulation sidebands, and delay echoes into a range the encoder features and the loss can represent. The resulting MSE gap between the two U-Nets is small (1.2 percentage points) but the FAD gap is larger (8.8 percentage points) and consistent (nine out of all effects), in line with input representation mattering for audio networks~\cite{cheuk}.

\begin{table}[t]
\centering
\caption{Mean pairwise L1 distance between predictions obtained by querying the trained model with all effect labels for a fixed clean input, divided by the mean L1 distance between each such prediction and the unprocessed input. A ratio near zero would indicate that predictions do not vary appreciably with the conditioning label. This table is reported as an informal check of the best U-Net model not collapsing.}
\label{tab:conditioning-sensitivity}
\small
\begin{tabular}{lr}
\toprule
Effect & Mean ratio \\
\midrule
\quad RAT             & 1.458 \\
\quad Tube Screamer   & 1.489 \\
\quad Flanger         & 1.463 \\
\quad Phaser          & 1.432 \\
\quad Digital Delay   & 1.431 \\
\quad Sweep Echo      & 1.422 \\
\quad Tape Echo       & 1.475 \\
\quad Hall Reverb     & 1.132 \\
\quad Plate Reverb    & 1.462 \\
\quad Spring Reverb   & 1.397 \\
\midrule
\multicolumn{2}{l}{\textit{Mean across effects}} \\[2pt]
\quad All effects     & 1.416 \\
\bottomrule
\end{tabular}
\end{table}

As an informal diagnostic, we examine whether the U-Net responds to the conditioning label rather than collapsing toward a single transformation. For each clean input, we generate outputs for all effect labels and compute the mean pairwise L1 distance between predictions, normalized by the mean L1 distance between each prediction and the unprocessed input (Table~\ref{tab:conditioning-sensitivity}). Ratios substantially above zero indicate label-dependent variation. All effects produce ratios above 1.0 (mean 1.416), providing qualitative evidence that the conditioning signal influences the generated transformations rather than being ignored by the model. This diagnostic should be interpreted only as evidence against conditioning collapse, not as validation of effect-specific generation.

The demo website probes out-of-distribution behavior, pairing four EGFxSet-derived passages with Guitar-TECHS~\cite{guitartechs} real performance recordings, each processed by the log-magnitude U-Net and the ConvVAE under all effects. On Guitar-TECHS input the U-Net produces audible but weaker transformations, and the ConvVAE shows the same smoothed-upper-band failure seen in-distribution. The demo therefore contrasts with the EGFxSet-derived distribution, within which the headline results hold.

\bibliographystyle{IEEEtranDAFx}
\bibliography{DAFx26_tmpl} 

\end{document}